\input harvmac

\overfullrule=0pt
\abovedisplayskip=12pt plus 3pt minus 1pt
\belowdisplayskip=14pt plus 3pt minus 1pt

%
\def\tilde{\widetilde}

\def\to{\rightarrow}
\def\tphi{{\tilde\phi}}

\def\tC{{\tilde C}}

\def\tF{{\tilde F}}

\def\bigone{\hbox{1\kern -.23em {\rm l}}}
\def\ZZ{\hbox{\zfont Z\kern-.4emZ}}
\def\half{{\litfont {1 \over 2}}}

\def\hA{{\hat A}}

\def\hF{{\hat F}}

\font\litfont=cmr6

\def\tX{{\tilde X}}

\def\tpa{{2\pi\alpha'}}

\def\ola{\overleftarrow}
\def\ora{\overrightarrow}
\def\tPsi{{\tilde\Psi}}
\def\cK{{\cal K}}
\font\smit=cmti8 scaled\magstep1

\lref\MukhiSS{
S.~Mukhi,
{\it ``Star products from commutative string theory''},
Pramana {\bf 58}, 21 (2002) [arXiv:hep-th/0108072].
}
\lref\seiwit{N. Seiberg and E. Witten, {\it ``String theory and
noncommutative geometry''}, 
JHEP {\bf 9909}, 032 (1999) [arXiv:hep-th/9908142].}
\lref\seibnew{N. Seiberg, {\it ``A note on background
independence in noncommutative gauge theories, matrix model and
tachyon condensation''}, 
JHEP {\bf 0009}, 003 (2000) [arXiv:hep-th/0008013].}
\lref\garousi{M. Garousi, {\it ``Noncommutative world volume
interactions on D-branes and Dirac-Born-Infeld action''},
Nucl. Phys. {\bf B579} 209 (2000) [arXiv:hep-th/9909214].}
\lref\wyllard{N. Wyllard, {\it ``Derivative corrections to D-brane 
actions with constant background  fields''}, 
Nucl. Phys. {\bf B598}, 247 (2001) [arXiv:hep-th/0008125].}
\lref\earlyderivcomp{
J.~H.~Schwarz, {\it ``Superstring Theory''}, Physics Reports {\bf 89},
223 (1982)\semi A.~A. Tseytlin, {\it ``Vector field effective actions
in the open superstring theory''}, Nucl. Phys. {\bf B276}, 391
(1986)\semi 
O.~D. Andreev and A.~A. Tseytlin, {\it ``Partition function
representation for the open superstring effective action: cancellation
of M\"obius infinities and derivative corrections to the Born-Infeld
lagrangian''}, Nucl. Phys. {\bf B311}, 205 (1988)\semi
K.~Hashimoto, {\it ``Corrections to D-brane action and generalized
boundary state''}, Phys. Rev. {\bf D61} 106002 (2000),
[arXiv:hep-th/9909027]\semi
K. Hashimoto, {\it ``Generalized supersymmetric boundary state''},
JHEP {\bf 0004}, 023 (2000) [arXiv:hep-th/9909095].}
\lref\earlyderivnc{L.~Cornalba and R.~Schiappa,
{\it ``Matrix theory star products from the Born-Infeld action''},
hep-th/9907211\semi
Y.~Okawa,
{\it ``derivative corrections to Dirac-Born-Infeld lagrangian and 
non-commutative gauge theory''}, hep-th/9909132, 
Nucl.\ Phys.\ {\bf B566}, 348 (2000)\semi
L.~Cornalba,
{\it ``Corrections to the Abelian Born-Infeld action arising from  
noncommutative geometry''}, hep-th/9912293, JHEP {\bf 0009}, 017
(2000)\semi
S.~Terashima,
{\it ``On the equivalence Between noncommutative and ordinary gauge 
theories''}, hep-th/0001111, JHEP {\bf 0002}, 029 (2000)\semi
Y.~Okawa and S.~Terashima,
{\it ``Constraints on effective lagrangian of D-branes from 
non-commutative gauge theory''}, hep-th/0002194, 
Nucl.\ Phys.\ {\bf B584}, 329 (2000)\semi
L.~Cornalba,
{\it ``On the general structure of the non-Abelian Born-Infeld
action''}, hep-th/0006018.}
\lref\dms{S.~R. Das, S. Mukhi and N.~V. Suryanarayana, {\it ``Derivative 
corrections from noncommutativity''}, JHEP {\bf 0108}, 039 (2001)
[arXiv:hep-th/0106024].} 
\lref\callan{C.~G. Callan, C.~Lovelace, C.~R.~Nappi 
and S.~A. Yost, {\it ``String loop   corrections to beta functions''},
Nucl. Phys. {\bf B288} 525 (1987)\semi
C.~G. Callan, C.~Lovelace, C.~R.~Nappi, and S.~A. Yost, 
{\it ``Adding holes and crosscaps to the superstring''},
Nucl.~Phys.~{\bf B293} 83 (1987)\semi
C.~G. Callan, C.~Lovelace, C.~R. Nappi, and S.~A. Yost, 
{\it ``Loop corrections to superstring equations of motion''}
Nucl. Phys. {\bf B308} (1988) 221.}
\lref\divecchia{P.~{Di Vecchia} and A.~Liccardo, 
{\it ``D branes in string theory, I''}, [arXiv:hep-th/9912161]\semi
P.~{Di Vecchia} and A.~Liccardo, 
{\it ``D branes in string theory, II''}, [arXiv:hep-th/9912275].}
\lref\MukhiZM{
S.~Mukhi and N.~V.~Suryanarayana,
{\it ``Chern-Simons terms on noncommutative branes''},
JHEP {\bf 0011}, 006 (2000) [arXiv:hep-th/0009101].
}
\lref\okawao{Y. Okawa and H. Ooguri, {\it ``An exact solution to 
Seiberg-Witten equation of noncommutative gauge theory''},
Phys.\ Rev.\ D {\bf 64}, 046009 (2001) [arXiv:hep-th/0104036].}
\lref\liu{H. Liu, {\it ``~$*$-Trek II: $*_n$ operations, open
Wilson lines and the Seiberg-Witten map''},
Nucl.\ Phys.\ B {\bf 614}, 305 (2001) [arXiv:hep-th/0011125].}
\lref\dastriv{S.~R. Das and S. Trivedi, {\it ``Supergravity
couplings to noncommutative branes, open Wilson lines and generalized
star products''}, JHEP {\bf 02} (2001) 046 [arXiv:hep-th/0011131].}
\lref\LiuPK{
H.~Liu and J.~Michelson,
{\it ``Ramond-Ramond couplings of noncommutative D-branes''},
Phys.\ Lett.\ B {\bf 518}, 143 (2001)
[arXiv:hep-th/0104139].
}
\lref\MukhiVX{
S.~Mukhi and N.~V.~Suryanarayana,
{\it ``Gauge-invariant couplings of noncommutative branes to Ramond-Ramond
backgrounds''},
JHEP {\bf 0105}, 023 (2001)
[arXiv:hep-th/0104045].
}
\lref\MukhiYS{
S.~Mukhi and N.~V.~Suryanarayana,
{\it ``Ramond-Ramond couplings of noncommutative branes''},
[arXiv:hep-th/0107087].
}
\lref\LiuQA{
H.~Liu and J.~Michelson,
{\it ``*-trek III: The search for Ramond-Ramond couplings''},
Nucl.\ Phys.\ B {\bf 614}, 330 (2001)
[arXiv:hep-th/0107172].
}
\lref\garousi{M. Garousi, {\it ``Noncommutative world volume
interactions on D-branes and Dirac-Born-Infeld action''},
hep-th/9909214, Nucl. Phys. {\bf B579} 209 (2000).}
\lref\starn{H. Liu and J. Michelson, {\it ``~$*$-Trek: the one
loop N=4 noncommutative SYM action''}, hep-th/0008205\semi
F.~Ardalan and N.~Sadooghi,
{\it ``anomaly and nonplanar diagrams in noncommutative gauge
theories''}, hep-th/0009233\semi
T. Mehen and M. Wise, {\it ``Generalized $*$-products, 
Wilson lines and the solution of the Seiberg-Witten equations''},
hep-th/0010204, JHEP {\bf 0012}, 008 (2000)\semi
A.~Santambrogio and D.~Zanon,
{\it ``One-loop four-point function in noncommutative N=4 Yang-Mills 
theory''}, hep-th/0010275, JHEP {\bf 0101}, 024 (2001)\semi
Y.~Kiem, D.~H.~Park and S.~Lee,
{\it ``Factorization and generalized *-products''},
hep-th/0011233, Phys.\ Rev.\ D {\bf 63}, 126006 (2001)\semi
K.~Okuyama,
{\it ``Comments on open Wilson lines and generalized star products''},
hep-th/0101177, Phys.\ Lett.\ B {\bf 506}, 377 (2001).}
\lref\WittenNZ{
E.~Witten,
{\it ``Noncommutative tachyons and string field theory''},
arXiv:hep-th/0006071.
}
\lref\kontsevich{M. Kontsevich, {\it ``Deformation quantization of
Poisson manifolds I''}, q-alg/9709040.}
\lref\CornalbaSM{
L.~Cornalba and R.~Schiappa,
{\it ``Nonassociative star product deformations for D-brane 
worldvolumes in  curved backgrounds''},
Commun.\ Math.\ Phys.\  {\bf 225}, 33 (2002)
[arXiv:hep-th/0101219].
}
\lref\iikk{N. Ishibashi, S. Iso, H. Kawai and Y. Kitazawa, {\it
``Wilson loops in noncommutative Yang-Mills''},
hep-th/9910004, Nucl. Phys. {\bf B573}, 573 (2000).
}
\lref\CornalbaAH{
L.~Cornalba,
{\it ``D-brane physics and noncommutative Yang-Mills theory''},
Adv.\ Theor.\ Math.\ Phys.\  {\bf 4}, 271 (2000)
[arXiv:hep-th/9909081].
}
\lref\IshibashiVI{
N.~Ishibashi,
{\it ``A relation between commutative and noncommutative descriptions of  
D-branes''},
arXiv:hep-th/9909176.
}
\lref\GrangeGC{
P.~Grange,
{\it ``Derivative corrections from boundary state computations''},
arXiv:hep-th/0207211.
}
\lref\PalXP{
S.~S.~Pal,
{\it ``Derivative corrections to Dirac-Born-Infeld and Chern-Simon 
actions  from non-commutativity''},
Int.\ J.\ Mod.\ Phys.\ A {\bf 17}, 1253 (2002)
[arXiv:hep-th/0108104].
}

{\nopagenumbers
\Title{\vbox{
\hbox{hep-th/0208203}
\hbox{TIFR/TH/02-28}
\hbox{DAMTP-2002-111}}}
{\vbox{
\centerline{Open-String Actions and Noncommutativity}
\medskip
\centerline{Beyond the Large-$B$ Limit}}}
\vskip -20pt

\centerline{Sunil Mukhi$^{a,b,}$\foot{On 
sabbatical leave from the Tata Institute
of Fundamental Research, Mumbai.} and Nemani V. Suryanarayana$^c$}
\vskip 8pt
\centerline{$^a$\smit School of Natural Sciences, Institute for 
Advanced Study}
\vskip -3pt
\centerline{\smit Princeton, NJ 08540, U.S.A.}
\medskip
\centerline{$^b$\smit Tata Institute of Fundamental Research}
\vskip -3pt
\centerline{\smit Mumbai 400 005, India}
\medskip
\centerline{$^c$\smit DAMTP, Centre for Mathematical Sciences}
\vskip -3pt
\centerline{\smit Cambridge CB3 0WA, U.K.}

\vskip 0.8truecm
\centerline{\bf ABSTRACT}
\smallskip
In the limit of large, constant $B$-field (the ``Seiberg-Witten
limit''), the derivative expansion for open-superstring effective
actions is naturally expressed in terms of the symmetric products
$*_n$. Here, we investigate corrections around the large-$B$ limit,
for Chern-Simons couplings on the brane and to quadratic order in gauge
fields. We perform a boundary-state computation in the commutative
theory, and compare it with the corresponding computation on the
noncommutative side. These results are then used to examine the
possible role of Wilson lines beyond the Seiberg-Witten limit. To
quadratic order in fields, the entire tree-level
amplitude is described by a metric-dependent deformation of the $*_2$
product, which can be interpreted in terms of a deformed
(non-associative) version of the Moyal $*$ product.

\vfill
\Date{August 2002}
\eject}

\ftno=0

\newsec{Introduction}
\medskip

In the background of a constant and large $B$-field, the structure of
higher-derivative terms in the open-superstring effective action is
encoded in certain symmetric products $*_n$, as follows from
Refs.\refs{\garousi,\liu,\dastriv,\MukhiVX,\okawao,\LiuPK,%
\dms,\MukhiSS}\foot{Here
and in the following, ``open-string actions'' will mean the tree-level
low-energy effective action for the coupling of open-string modes to
linearized closed-string modes. We will always take the closed-string
fields to be at arbitrary nonzero momentum. We will work with a single
Euclidean D9-brane with a constant $B$-field of maximal rank. The
extension to D-branes of lower dimension is straightforward. However,
the nonabelian case involving multiple D-branes seems much more
difficult, and is beyond the scope of this paper.}. The
higher-derivative terms that have been understood in this way are
those which are leading at large $B$. This corresponds to the the
Seiberg-Witten (SW) limit\refs\seiwit, which is effectively the same
as the limit of very large $B$-field\refs\WittenNZ.  In the present
work, we will extend this understanding to terms that are subleading
at large $B$-field, or equivalently to corrections beyond the
Seiberg-Witten limit.

This approach to derivative corrections originally grew out of the
comparison between open-string actions in commutative and
noncommutative variables. The $*_2$ product was discovered in
Ref.\refs\garousi, and was generalized to $*_n$ and extensively
studied in Refs.\refs{\starn,\liu,\dastriv}. Following this,
noncommutative Ramond-Ramond couplings were studied, and comparison
with commutative Ramond-Ramond couplings\refs{\MukhiVX,\okawao,\LiuPK}
led to new topological identities on noncommutative gauge fields, as
well as a derivation of an explicit expression for the Seiberg-Witten
map\refs\seiwit, previously conjectured in Ref.\refs\liu.

Independently, higher derivative corrections to the commutative
open-superstring effective action (including couplings to both RR and
NS-NS closed string fields) were computed in Ref.\refs\wyllard\ using
the boundary-state formalism. These computations led to elegant
expressions that suggested an underlying mathematical structure. This
structure was later revealed\refs\dms\ to arise from the $*_n$
product, via the relation between noncommutative and commutative
actions. Some of these computations were extended to higher orders in 
Ref.\refs\PalXP.

Both commutative and noncommutative actions are infinite power series
in spacetime derivatives. Since derivatives carry a spatial dimension,
this has to be cancelled by a suitable dependence on the available
dimensional constants, which are $\alpha'$ and $B$. However, the two
actions organize these series differently. In particular, the
dependence of the commutative action on $B$ arises through the
generalized inverse metric
\eqn\invmet{
h^{ij}\equiv \left({1\over g+\tpa B}\right)^{ij} }
while in the noncommutative action the dependence arises through two
kinds of sources. One is the noncommutativity parameter $\theta$
arising in the Moyal $*$-product:
\eqn\moyal{
f(x) * g(x) \equiv f(x)\, e^{{i\over 2}\ola{~\partial_p}
\,\textstyle\theta^{pq}\ora{\,\partial_q\,}} g(x) }
while the other is the dependence of the action on the open-string
metric $G_{ij}$ and coupling $G_s$. 

The expressions for $\theta$ and $G$ in terms of closed-string
quantities depend on an antisymmetric matrix called the ``description
parameter''\refs{\seiwit,\seibnew}, denoted $\Phi_{ij}$, and we have:
\eqn\combin{
\left({1\over G+\tpa \Phi}\right)^{ij} + {\theta^{ij}\over \tpa}
= \left({1\over g+\tpa B}\right)^{ij}
}
The natural low-energy limit $\alpha'\to 0$ for open strings is called
the Seiberg-Witten (SW) limit. In this limit, $G,B,\Phi$ are kept fixed. It
follows from the above equation that $g_{ij}\sim\alpha'^2$ and so the
closed-string metric goes to zero faster than $\alpha' B$. In this
sense it is a large-$B$ limit.

Expanding about this limit, we find that:
\eqn\expandsw{
\eqalign{
\theta^{ij} &= \left({1\over B}\right)^{ij} + {\cal O}(\alpha')\cr
G^{ij} &= -{\theta^{ik}g_{kl}\theta^{lj}\over(\tpa)^2} + 
{\cal O}(\alpha')\cr}}
Note that the leading terms do not depend on the description parameter
$\Phi$, only the subleading corrections depend on it. For a particular
value $\Phi=-B$, the corrections in the above equation vanish to all
orders.

To compute derivative corrections, to high orders in derivatives, has
traditionally been a difficult technical exercise\foot{Earlier work on
derivative corrections in open-string theory can be found in
Ref.\refs\earlyderivcomp.}. 
It turns out that noncommutativity is a
handy tool to obtain large parts of the result. In the limit of large
$B$, one makes the replacement:
\eqn\hrepl{
h^{ij} \to {\theta^{ij}\over \tpa} }
On the commutative side this gives rise to infinitely many derivative
corrections that, after the above replacement, are all of lowest order
in $\alpha'$ -- though they originally arose from terms of arbitrary
order in $\alpha'$. On the noncommutative side, this limit suppresses
all but the leading zero-derivative term (DBI or Chern-Simons), and
the dependence on $\theta$ is encoded in the Moyal product.  The
closed-string modes in the action are taken to have arbitrary nonzero
momentum (otherwise the terms we are computing would all vanish upon
partial integration). Hence one has to make use of the open Wilson
line\refs\iikk\ prescription to get a gauge invariant result, leading
to additional dependence on $\theta$. The result, described in detail
in Ref.\refs\dms, is that the derivative expansion at large $B$ is
encoded in the $*_n$ products\foot{Previous attempts to constrain
effective actions using noncommutativity can be found in
Ref.\refs\earlyderivnc.}.

As a check, predictions from noncommutativity were compared to the
existing perturbative open-string amplitudes in the literature.
Striking agreement was found between these predictions and the
coefficients and tensor structures computed explicitly in 
Ref.\refs\wyllard\ to low orders in derivatives.

Inspired by this agreement and the elegant predictions of
noncommutativity, a new perturbative amplitude calculation was
performed in Ref.\MukhiSS. Here noncommutativity was not invoked, but
a specific class of derivative corrections was computed in the
commutative framework, to leading order in large $B$ and all orders in
derivatives. This was the correction to the Chern-Simons (CS) coupling
\eqn\cscoup{
S_{CS}^{(6)} = \half\int C^{(6)}\wedge F\wedge F }
where $C^{(6)}$ is the Ramond-Ramond 6-form potential. (Recently, this
calculation was extended to other couplings on the 
D-brane worldvolume\refs\GrangeGC.)

In the Seiberg-Witten limit ($\alpha'\to 0$ with $g\sim \alpha'^2$) it
was shown by an explicit boundary-state computation that the above
coupling is corrected to:
\eqn\cscoupcorr{
S_{CS}^{(6)} = \half\int C^{(6)}\wedge \langle F\wedge F \rangle_{*_2}}
where
\eqn\startwodef{
\langle F_{ij}(x),F_{kl}(x) \rangle_{*_2} \equiv 
F_{ij}(x){\sin(\half\ola{~\partial_p}\,\theta^{pq}\ora{\,\partial_q\,})
\over \half\ola{~\partial_p}\,\theta^{pq}\ora{\,\partial_q\,} } 
F_{kl}(x) }
This result holds to quadratic order in $F$, with higher-order
corrections. The above expression agrees perfectly with a prediction
from noncommutativity that was made in Ref.\refs\dms. This prediction
and the explicit computation which it confirms are reviewed in Section
2, to establish the techniques and notation for the subsequent
sections.

One striking feature of Eq.\cscoupcorr\ is that it describes a set of
derivative corrections all of which have rational numerical
coefficients, which come from expanding the definition of $*_2$
above. Indeed, going beyond the specific CS coupling above, it follows
from Ref.\refs\dms\ that the entire leading behaviour of open-string
theory at large $B$-field is given by terms with rational coefficients
that come from the expansion of $*_n$ products along with
determinants, denominators and Wilson lines. This means that the
transcendental coefficients of string theory, such as $\zeta(2n+1)$,
must all drop out in this limit\foot{We thank Ashoke Sen for stressing
this point.}. It would be interesting to find an independent reason
for this.

Thus it is tempting to ask what happens if we go beyond the
Seiberg-Witten approximation.  For example, in the commutative theory,
to first order in $\alpha'$ one would make the replacement:
\eqn\hrepfirst{
h^{ij}\equiv \left({1\over g+\tpa B}\right)^{ij}
~\sim~ {\theta^{ij}\over\tpa} - {(\theta g\theta)^{ij}\over
(\tpa)^2} }
It is important to keep in mind that the second term is of order
$\alpha'$ relative to the first one, because of the scaling of
$g\sim\alpha'^2$ implicit in the Seiberg-Witten limit. We can also
write:
\eqn\alsowrite{
h^{ij}\equiv \left({1\over g+\tpa B}\right)^{ij}~\sim~
{\theta^{ij}\over\tpa} + G^{ij} }
From Eq.\combin\ we see that this equation is exact in the $\Phi=0$
description.

In first subleading order, it is reasonable to expect that
transcendental coefficients reappear in the computation of
amplitudes. This would mean that we are no longer working with an
excessively over-simplified limit of open-string theory. It turns out
that the perturbative calculation of Ref.\MukhiSS\ which led to the
result in Eq.\cscoupcorr\ can be extended without too much trouble to
order $\alpha'$, and we will describe this computation below.

An exact tree-level computation of the three-point function of two
gauge fields with a Ramond-Ramond potential in noncommutative string
theory was performed in Ref.\refs\LiuQA. This can be expanded to order
$\alpha'$ and compared with the above result. The two computations
will be seen to agree perfectly.

The paper is organized as follows. In Section 2, we review the
boundary-state computation of open-string actions, and the computation
of the CS coupling to leading order in the large-$B$ limit performed
in Ref.\refs\MukhiSS. All the computations discussed here and
subsequently are carried out in an approximation where we restrict to
quadratic order in gauge potentials, but all orders in derivatives.
In Section 3 we use the boundary-state formalism to compute the first
correction around the large-$B$ limit. In Section 4 this result is
compared with a computation of Liu and Michelson\refs\LiuQA\ of the
corresponding amplitude on the noncommutative side. In Section 5, we
use the computation of Ref.\refs\LiuQA\ to demonstrate that the entire
tree-level amplitude involving two open-string states and one RR
potential is encoded in a deformed $*_2$ product. This in turn can be
seen to arise from a deformed (non-associative) version of the Moyal
$*$ product, and a standard prescription involving straight open
Wilson lines. Finally, in an Appendix we give direct world-sheet
derivations of the proposals made in Section 5.

\newsec{Predictions from Noncommutativity and
Boundary-State Computations}

This section is a review of Refs.\refs{\dms,\wyllard,\MukhiSS}. Let us
focus on a particular Chern-Simons coupling, the one
involving the Ramond-Ramond 6-form $C^{(6)}$. In the commutative
theory, this is just
\eqn\commcs{
\half\int C^{(6)}\wedge F\wedge F}
As shown in Ref.\refs\MukhiZM, the noncommutative version of this
coupling is obtained from the
commutative one by making the replacement:
\eqn\ncrepl{
F\to \hF{1\over 1-\theta\hF} }
where $\hF$ is the noncommutative gauge field strength,
multiplying the action by a factor $\sqrt{\det(1-\theta\hF)}$, and
using the Moyal $*$-product defined in Eq.\moyal.

Finally, to make a coupling that is gauge-invariant even for
nonconstant fields, this has to be combined with an open Wilson
line\refs{\liu,\dastriv}. The resulting
expression\refs{\MukhiVX,\okawao,\LiuPK\MukhiYS}\ for the coupling to
$C^{(6)}$ is more conveniently expressed in momentum space, where
${\tilde C}^{(6)}(k)$ is the Fourier transform of $C^{(6)}(x)$:
\eqn\csthree{
\half {\tilde C}^{(6)}(-k)\wedge
\int L_*\Big[{\sqrt{\det(1-\theta \hat F)}}
\left(\hF{1\over 1-\theta\hF}\right)\wedge\,
\left(\hF{1\over 1-\theta\hF}\right) W(x, C)\Big] * e^{ik.x}}
Here $W(x,C)$ is an open Wilson line, and $L_*$ is the prescription of
smearing local operators along the Wilson line and path-ordering with
respect to the Moyal product. Evaluation of the $L_*$ prescription
leads to $*_n$ products\refs{\liu,\dastriv}. While expressions such as
the above were originally obtained in the $\Phi=-B$ description, it
has been argued (see Refs.\refs{\dastriv,\MukhiYS,\LiuPK}) that they
actually hold in all descriptions, provided the value of $\theta$
appropriate to the given description is used in the $*$ products.

In the approximation where we retain only terms that
are quadratic in $F$, one can ignore the Wilson line, the
denominators, the Pfaffian prefactor and the Seiberg-Witten map
relating $\hF$ to $F$. Then the above expression can easily be
re-expressed in position space, and it turns into:
\eqn\quadapprox{
\half\int C^{(6)}\wedge \langle F\wedge F\rangle_{*_2} }

Let us now see how this result can be obtained directly by a
computation in commutative string theory. The computation will be
done in the boundary-state formalism. Some background on how to
compute derivative corrections in this formalism may be found in
Ref.\refs\wyllard, and we will largely follow the notation in that
paper. 

Let us denote the sum of all derivative corrections to $S_{CS}$ as
$\Delta S_{CS}$. Our starting point is the expression
\eqn\wylstart{
S_{CS} + \Delta S_{CS} =
\big\langle C \big| e^{-{i\over \tpa} \int d\sigma d\theta D\phi^i
A_i(\phi) } \big|B\big\rangle_{R} }
where $|C\rangle$ represents the RR field, and $|B\rangle_R$ is the
Ramond-sector boundary state for zero field strength. We are using
superspace notation, for example $\phi^i = X^i + \theta \psi^i$
and $D$ is the supercovariant derivative.

Combining Eqs.(2.3),(2.6),(2.13) of Ref.\refs\wyllard, we can
rewrite this as:
\eqn\wylnext{
\eqalign{
S_{CS} + \Delta S_{CS} &=
\big\langle C\big| 
e^{{i\over \tpa} \int d\sigma d\theta\sum_{k=0}^\infty{1\over (k+1)!}
{k+1\over k+2} D\tphi^j\tphi^i \tphi^{a_1}\cdots
\tphi^{a_k}\del_{a_1}\ldots\del_{a_k}
F_{ij}(x)}\times\cr
&\phantom{\langle C|}
e^{{i\over\tpa} \int d\sigma[\tPsi^i \psi_0^j
+ \psi_0^i \psi_0^j]\sum_{k=0}^\infty{1\over k!}\tX^{a_1}
\cdots \tX^{a_k} \del_{a_1}\ldots\del_{a_k} 
F_{ij}(x)}\big|B\big\rangle_{R}\cr} }
where nonzero modes have a tilde on them, while the zero modes are
explicitly indicated.

Since we are looking for couplings to the RR 6-form $C^{(6)}$, and
working to order $F^2$, we only need terms with the structure
$\del\ldots\del F\wedge\del\ldots\del F$. For such terms, two $F$'s
and 4 $\psi_0$'s must be retained. Thus we can drop the first
exponential factor in Eq.\wylnext\ above, as well as the first fermion
bilinear $\tPsi^i \psi_0^j$ in the second exponential. Then,
expanding the exponential to second order, we get:
\eqn\exposecond{
\eqalign{
S_{CS} + \Delta S_{CS} =& ~
\half\sum_{n=0}^\infty \sum_{p=0}^\infty \left({i\over\tpa}\right)^2
\int_0^{2\pi} \!\! d\sigma_1\int_0^{2\pi} \!\! d\sigma_2 ~
\big\langle C\big| \left(\half\psi_0^i\psi_0^j\right)
\left(\half\psi_0^k\psi_0^l\right)~\times\cr
& {1\over n!} \tX^{a_1}(\sigma_1) \cdots \tX^{a_n}(\sigma_1)
{1\over p!} \tX^{b_1}(\sigma_2) \cdots \tX^{b_p}(\sigma_2)~\times\cr
&\del_{a_1}\ldots\del_{a_n} F_{ij}(x)\,
\del_{b_1}\ldots\del_{b_p} F_{kl}(x)
\big|B\big\rangle_{R} \cr}}
Now we need to evaluate the 2-point functions of the $\tX$. The
relevant contributions have non-logarithmic finite parts\refs\wyllard\
and come from propagators for which there is no self-contraction. This
requires that $n=p$. Then we get a combinatorial factor of $n!$ from
the number of such contractions in $\big\langle
\big(\tX(\sigma_1)\big)^n\,
\big(\tX(\sigma_2)\big)^n\big\rangle$. The result is:
\eqn\contrac{
\eqalign{
S_{CS} + \Delta S_{CS} =& ~
 \half
\sum_{n=0}^\infty
{1\over n!} \left({i\over\tpa}\right)^2
\int_0^{2\pi} \!\! d\sigma_1\int_0^{2\pi} \!\! d\sigma_2 ~
D^{a_1 b_1}(\sigma_1-\sigma_2)\cdots 
D^{a_n b_n}(\sigma_1-\sigma_2)~\times\cr
&\del_{a_1}\ldots\del_{a_n} F_{ij}(x)\,
\del_{ b_1}\ldots\del_{b_n} F_{kl}(x)
~\big\langle C\big|\left(\half\psi_0^i\psi_0^j\right)
\left(\half\psi_0^k\psi_0^l\right)\big|B\big\rangle_{R}\cr }}
The fermion zero mode expectation values are evaluated using the
recipe:
\eqn\zerorecipe{
\half \psi_0^i\psi_0^j F_{ij} \rightarrow
(-i\alpha')F }
where the $F$ on the right hand side is a differential 2-form.
The justification for this can be found below Eq.(B.3) of 
Ref.\refs\wyllard. Thus we are led to:
\eqn\fermizero{
S_{CS} + \Delta S_{CS} =
T^{a_1\ldots a_n;\,b_1\ldots b_n} 
\,\del_{a_1}\ldots\del_{a_n} F \wedge
\del_{b_1}\ldots\del_{b_n} F}
where
\eqn\ttensor{
T^{a_1\ldots a_n;\,b_1\ldots b_n} \equiv~ 
\half {1\over n!}\left({i\over\tpa}\right)^2 (-i\alpha')^2  
\int_0^{2\pi} \!\! d\sigma_1\int_0^{2\pi} \!\! d\sigma_2 \,
D^{a_1 b_1}(\sigma_1-\sigma_2)\cdots 
D^{a_n b_n}(\sigma_1-\sigma_2)}

Next we insert the expression for the propagator:
\eqn\propagat{
D^{ab}(\sigma_1 - \sigma_2)
= \alpha'\sum_{m=1}^\infty {e^{-\epsilon m}\over m}
\left(h^{ab}e^{im(\sigma_2-\sigma_1)}
+ h^{ba}e^{-im(\sigma_2-\sigma_1)} \right)}
where $\epsilon$ is a regulator, and
\eqn\hmunu{
h^{ij} \equiv {1\over g + \tpa B} }
As we have seen, this tensor when expanded about large $B$ has the
form:
\eqn\hexpand{
h^{ij}\sim 
{\theta^{ij}\over\tpa} - {(\theta g\theta)^{ij}\over
\tpa^2}  +\ldots }
where the terms in the expansion are alternately antisymmetric and
symmetric, and the two terms exhibited above are
description-independent. In this section we neglect all but the first
term above.

It follows that
\eqn\evalu{
T^{ a_1\ldots a_n;\, b_1\ldots  b_n}
=\half {1\over n!} (\alpha')^n 
\int_0^{2\pi} {d\sigma\over 2\pi}~
\prod_{i=1}^n \left(h^{a_i b_i} \sum_{m=1}^\infty
{e^{-\epsilon m + im\sigma}\over m}
+ h^{b_i a_i}\sum_{m=1}^\infty 
{e^{-\epsilon m -i m \sigma}\over m} \right) }
After evaluating the sum over $m$, the result, depending on the
regulator $\epsilon$, is
\eqn\evalusummed{
T^{ a_1\ldots a_n;\, b_1\ldots b_n}
=\half {1\over n!} (\alpha')^n
\int_0^{2\pi} {d\sigma\over 2\pi}~
\prod_{i=1}^n ~\Bigg(
- h^{[a_i b_i]} 
\ln {(1-e^{-\epsilon+ i\sigma})\over 
(1-e^{-\epsilon - i\sigma})}
 - h^{(a_i b_i)} 
\ln |1-e^{-\epsilon+ i\sigma}|^2
\Bigg) }
Here, $h^{[ab]}$ and $h^{(ab)}$ are, respectively, the antisymmetric
and symmetric parts of $h^{ab}$.

The large-$B$ or Seiberg-Witten limit consists of the replacements:
\eqn\swlimit{
h^{[ab]}\to {\theta^{ab}\over\tpa},\qquad h^{(ab)}\to 0 }
By virtue of the fact that
\eqn\logeval{
\lim_{\epsilon\to 0}~
\ln {(1-e^{-\epsilon+ i\sigma})\over 
(1-e^{-\epsilon - i\sigma})} = i(\sigma-\pi) }
this limit leads to an elementary integral. Evaluating it, one finally
obtains\refs\MukhiSS\ the result:
\eqn\derivsw{
\eqalign{
S_{CS} &+ \Delta S_{CS}\cr &= \half\int C^{(6)}\wedge \sum_{j=0}^\infty
(-1)^j {1\over 2^{2j} (2j+1)!} \,
\theta^{ a_1 b_1}\ldots \theta^{a_{2j} b_{2j}}\,
\del_{ a_1}\ldots\del_{a_{2j}} F \wedge
\del_{ b_1}\ldots\del_{b_{2j}} F \cr
&= 
\half\int C^{(6)}\wedge \langle F\wedge F\rangle_{*_2} }}
where the product $*_2$ was defined in Eq.\startwodef.
This agrees perfectly with Eq.\quadapprox, the prediction from
noncommutativity.

\newsec{Effective Action Beyond the Seiberg-Witten  Limit}

In this section, we extend the calculation from the point of
Eq.\evalusummed. To go to first order beyond the Seiberg-Witten limit,
we make the replacements:
\eqn\beyondsw{
h^{[ab]}\to {\theta^{ab}\over\tpa},\qquad h^{(ab)}\to -{\theta^{ac}
g_{cd}\theta^{db}\over (\tpa)^2} }
and keep all terms that are first order in $h^{(ab)}$.

Denote by $T_{(1)}^{ a_1\ldots a_n;\, b_1\ldots b_n}$ the first
correction to $T$ (defined in Eq.\ttensor)
away from the Seiberg-Witten limit. Then, we see that 
\eqn\toneexp{
T_{(1)}^{ a_1\ldots a_n;\, b_1\ldots b_n} = {(-i)^{n-1}\over 2}
{1\over (n-1)!} {1\over (2\pi)^n}
{(\theta g\theta)^{a_1 b_1}\over\tpa} \theta^{a_2 b_2}
\cdots \theta^{a_n b_n} \int_0^{2\pi} {d\sigma\over 2\pi}
(\sigma-\pi)^{n-1} \ln|1-e^{i\sigma}|^2 }
The integral in the above expression vanishes for even $n$. For odd
$n=2p+1$, we find that
\eqn\toneeval{
T_{(1)}^{ a_1\ldots a_{2p+1};\, b_1\ldots b_{2p+1}} =  {(-1)^p\over 2}
{1\over (2p)!} {1\over (2\pi)^{2p+1}}
{(\theta g\theta)^{a_1 b_1}\over\tpa} \theta^{a_2 b_2}
\cdots \theta^{a_{2p+1} b_{2p+1}} I_{2p+1}}
where 
\eqn\intdef{
\eqalign{
I_{2p+1}&\equiv \int_0^{2\pi} {d\sigma\over 2\pi}
(\sigma-\pi)^{2p} \ln|1-e^{i\sigma}|^2 \cr
&= 2 (-1)^p (2p)! \sum_{j=0}^{p-1} (-1)^j {\pi^{2j}\over
(2j+1)!} \zeta(2p-2j+1) }}

It is convenient to define a 4-form $W_4$ that encodes the 
derivative corrections for the coupling to $C^{(6)}$:
\eqn\wfourdef{
S_{CS}+\Delta S_{CS} = \half\int C^{(6)}\wedge F\wedge F
+ \int C^{(6)}\wedge W_4 }
As we have seen in the previous section, the leading-order term in
$W_4$ in the large-$B$ limit is:
\eqn\wfourleading{
W_4^{(0)} = \langle \,F\wedge F\,\rangle_{*_2}- F\wedge F }

The calculations leading to Eq.\intdef\ amount to computing
$W_4$ to first order (in $\alpha'$) around the Seiberg-Witten limit:
\eqn\derivcorrone{
\eqalign{
W_4^{(1)} &= \sum_{p=1}^\infty 
T_{(1)}^{ a_1\ldots a_{2p+1};\, b_1\ldots b_{2p+1}}
\partial_{a_1}\ldots \partial_{a_{2p+1}} F \wedge
\partial_{b_1}\ldots \partial_{b_{2p+1}} F\cr  &=
\sum_{p=1}^\infty {1\over (2\pi)^{2p+1}}
{(\theta g\theta)^{a_1 b_1}\over\tpa}
\theta^{a_2 b_2}\cdots \theta^{a_{2p+1} b_{2p+1}}
\partial_{a_1}\ldots \partial_{a_{2p+1}} F \wedge
\partial_{b_1}\ldots \partial_{b_{2p+1}} F~\times\cr
&\qquad\qquad\sum_{j=0}^{p-1} (-1)^j {\pi^{2j}\over (2j+1)!}
\zeta(2p-2j+1) \cr}}

Interchanging the order of the two summations, we find that the sum
over $j$ can be performed and leads to the appearance of the familiar
$*_2$ product. The result, after some relabelling of indices, is:
\eqn\derivcorrtwo{
W_4^{(1)} = \sum_{p=0}^\infty {\zeta(2p+3)\over (2\pi)^{2p+3}} 
{(\theta g\theta)^{cd}\over\tpa}
\theta^{a_1 b_1}\cdots \theta^{a_{2p+2} b_{2p+2}}~\times
\langle\, 
\partial_c \partial_{a_1}\ldots \partial_{a_{2p+2}} F\wedge\,
\partial_d \partial_{b_1}\ldots \partial_{b_{2p+2}} F\, \rangle_{*_2} }
Unlike the leading term Eq.\derivsw, which is a single infinite series
in derivatives summarised by the $*_2$ product, here we see a double
infinite series. After forming the $*_2$ product we still have an
additional series whose coefficients are $\zeta$-functions
of odd argument.

A more elegant representation of the result can be found by replacing
the $\zeta$-functions by their series representation: $\zeta(2p+3) =
\sum_{m=1}^\infty m^{-2p-3}$, and interchanging the $m$ and $p$
summations. This leads to:
\eqn\polerep{
W_4^{(1)} = {(\theta g\theta)^{cd}\over\tpa}
\langle\, \partial_c F\, \sum_{m=1}^\infty {1\over 2\pi m}
\Bigg({1\over 1- \Big({\ola{\partial_p\,} \textstyle\theta^{pq} 
\ora{\,\partial_q}\over \textstyle 2\pi
m}\Big)^2} -1 \Bigg)\wedge \partial_d F\,\rangle_{*_2} }
If the momenta of the two gauge fields are $k^{(1)}$ and $k^{(2)}$,
then we see that the term inside brackets
develops a pole whenever
\eqn\poleeqn{
k^{(1)}_p\, \theta^{pq}\, k^{(2)}_q = 2\pi m }
for any positive or negative integer $m$. Nevertheless, $W_4^{(1)}$,
and hence the effective action, remains nonsingular,
because at precisely the above values of momenta, the $*_2$ product
develops zeroes. 

Thus we have a finite expression for the first
correction to the effective action beyond the large-B limit. In the
following section we will compare this with a direct computation using
noncommutativity. 

\newsec{Comparison with a Noncommutative Amplitude Calculation}

The tree-level amplitude with two open-string vertex operators and one
closed string vertex operator on a disk has been evaluated exactly by
Liu and Michelson\refs\LiuQA. This calculation was performed in the
$\Phi=0$ description, which is most convenient for worldsheet
computations. In this section we are going to use it to compare with
results in the previous section that are of first order in
$\alpha'$. As we have noted, to this order the description is
irrelevant.

We will now use this to extract the term corresponding to our
computation in Eq.\derivcorrtwo\ and compare the two expressions. The
computation of Ref.\refs\LiuQA\ is an evaluation of the amplitude:
\eqn\twooneamp{
{\cal A}_2 \equiv \int_{-\infty}^\infty dy\,\left\langle 
V_{RR}^{-{1\over 2},-{3\over 2}}(q;i)\, V_O^0(a_1,k_1;0) \,
V_O^0(a_2,k_2;y)\right\rangle}
where $V_{RR}^{-{1\over 2},-{3\over 2}}$ is the vertex operator for an
RR potential of momentum $q$, in the $(-{1\over 2},-{3\over 2})$
picture, and $V_O^0$ are vertex operators for massless gauge fields of
momentum $k_i$ and polarizations $a_i$, $i=1,2$.

We define:
\eqn\tadef{
\eqalign{
t~ &\equiv~ \alpha' k_1\cdot k_2 ~=~ 
\alpha' {k_1}_i\, G^{ij}\, {k_2}_j \cr
a~&\equiv~ {1\over 2\pi}k_1\times k_2 ~=~
{1\over 2\pi} {k_1}_i\, \theta^{ij}\, {k_2}_j}}
and change integration variables via $y=-\cot\pi \tau$.
Then it follows (for details, see Sec.(3.2) and Appendix D of 
Ref.\refs\LiuQA) that the coefficient of $F\wedge F$ provided by this
computation, to be compared with the coefficient of Eq.\derivcorrtwo, is:
\eqn\ifourint{
2^{2t}\int_{0}^{\half} d\tau\, (\cos\pi\tau)^{2t}\,
\cos 2\pi a \tau= \half
{{\Gamma(1+2t)}\over{\Gamma(1+a+t)\Gamma(1-a+t)}}}

The RHS can be expanded in powers of $t$ and, up to terms of ${\cal
O}(t^2)$, one has:
\eqn\gammaexp{
{{\Gamma(1+2t)}\over{\Gamma(1+a+t)\Gamma(1-a+t)}} =
{1\over{\Gamma(1-a)\Gamma(1+a)}}\Big[1-\left(2\gamma
+\psi(1-a)+\psi(1+a)\right)t + {\cal O}(t^2)\Big]}
where $\gamma$ is the Euler constant and $\psi(x)$ is the digamma
function ${d \over {dx}}\ln\Gamma(x)$. 

The first term can be recognised as the kernel of the $*_2$-product,
using the relation:
\eqn\starker{
{1\over{\Gamma(1-a)\Gamma(1+a)}} = {\sin\pi a\over{\pi a}}}

Let us now examine the second term more carefully.
We use the fact that:
\eqn\nextterm{
\psi(1+x) = -\gamma + \sum_{k=2}^{\infty}(-1)^k \zeta(k) x^{k-1}}
to write: 
\eqn\zetaseries{
\eqalign{
2\gamma +\psi(1-a)+\psi(1+a) &=
-\sum_{k=2}^{\infty}\left(1-(-1)^k\right)\zeta(k)\,a^{k-1}\cr
&= -2\sum_{p=0}^{\infty}\zeta(2p+3)\,a^{2p+2}}}
Putting everything together, we find that:
\eqn\finalcorr{
W_4^{(1)} =
{{\sin\left({{k_1\times k_2}\over 2}\right)}\over{{k_1\times k_2}\over
2}}\,\sum_{p=0}^{\infty}{{\zeta(2p+3)}
\over{\tpa(2\pi)^{2p+3}}}\,(k_1\times k_2)^{2p+2}\,
\Big({k_1}_i(\theta g \theta)^{ij}{k_2}_j\Big) \tF(k_1)\wedge\tF(k_2)}
On Fourier transforming, this is identical to Eq.\derivcorrtwo\ of the
previous section.

\newsec{Wilson Lines and Deformed $*$-products}

We have seen in the previous section that the correction to the
commutative Chern-Simons action found in section 3 matches with a
corresponding computation in noncommutative language. On general
grounds, one expects that the commutative and the noncommutative
actions match precisely to all orders upon using the SW map. Therefore
it is natural to ask how the correction computed in this paper can be
re-expressed as an effective action in terms of the noncommutative
field variables. 

As discussed in Section 2, the leading term for the Chern-Simons
coupling in the Seiberg-Witten limit is given in terms of straight
Wilson lines by Eq.\csthree. In what follows, we attempt to interpret
the corrections beyond the SW limit  in the language of open Wilson
lines. In subsections (5.1) and (5.2), we investigate the consequences
for the topological identity of Refs.\refs{\MukhiVX,\okawao,\LiuPK}\
and the Seiberg-Witten map beyond the SW limit. We restrict ourselves
to quadratic order in gauge fields. This enables us to make use of the
explicit computation in Ref.\refs\LiuQA, and also to avoid many
technical complications. In order to get a complete picture one will
have to go beyond the quadratic approximation, an issue which we
intend to address in future work.

As we are now going to use the results of Ref.\refs\LiuQA\ to all
orders in the expansion around the SW limit, it will be important to
work with a specific description, namely $\Phi=0$, the one in which
these calculations were performed. Henceforth it will be understood
that we are always in this description.

Our first observation will be that the corrections away from the SW
limit can be reproduced by modifying the prescription for the
noncommutative effective action. The usual prescription, valid in the
SW limit, is to smear local operators along an open Wilson line that
runs along a specific straight contour. We will show that corrections
can be incorporated by modifying the way in which local operators are
smeared along the contour. The modification consists of inserting a
specific function of the smearing parameter, and leads to a deformed
$*_2$ product.  Next we will find an equivalent but more suggestive
way to incorporate the corrections. This uses the standard smearing
prescription and Wilson line, but the Moyal $*$ product is deformed in
a particular way. The relationship between deformed $*_n$ products and
deformed Moyal $*$-products will be seen to arise in many contexts,
which we take as an encouraging indication that this extension of the
usual picture is a natural one. We will be led to conjecture that the
entire large-B expansion on the noncommutative side is encoded in a
deformed $*$-product or series of such products, along with the
prescription of smearing over a straight Wilson line.

Let us start by reviewing how the $*_2$ product originates
from smearing. The straight Wilson line runs along a contour:
\eqn\strtcont{
x^i(\tau) = x^i + \theta^{ij}k_j\tau}
where $\tau \in [0,1]$. As long as we work to quadratic order in the
commutative field strength $F$, we can replace ${\hat F}$ by $F$ in
Eq.\csthree. The only effect of the Wilson line in this case is that
the operators are path-ordered and smeared along its contour.  One of
the inserted operators may be fixed to lie at the starting point of
the contour \strtcont. Thus for a product of two local operators, we
only need to smear the second one linearly over the contour. In
quadratic order, Eq.\csthree\ therefore reduces to:
\eqn\csquad{
\half\tC^{(6)}(-k)\wedge
\int_0^1 d\tau\int dx~ F(x)*\wedge\, F(x+\theta\cdot k \tau)\,
e^{ik.x}} 
From now on we will drop the $\tC^{(6)}$ factor for simplicity.

Going to momentum space\foot{The integration measures $dx$, $dk_i$ are
defined to implicitly include the necessary factors of $2\pi$.}, 
we have:
\eqn\csquadmom{
\half \int_0^1 d\tau\int dx~ dk_1~dk_2~
e^{-{i\over 2}k_1\times k_2}
\tF(k_1)\wedge\tF(k_2)~e^{-ik_1.x}e^{-ik_2.(x+\theta\cdot k\tau)}
e^{ik.x}}
The $\tau$ and $x$ integrals are easily evaluated, leading to:
\eqn\cstaueval{
\half \int dk_1\,dk_2~{\sin{k_1\times k_2\over
2}\over {k_1\times k_2\over 2}}\,
\tF(k_1)\wedge\tF(k_2)\, \delta(k_1+k_2-k)}
This is the result that we have quoted several times in previous
sections, since
\eqn\csstartwo{
{\sin{k_1\times k_2\over
2}\over {k_1\times k_2\over 2}}\,
\tF(k_1)\wedge\tF(k_2)\equiv \langle\tF(k_1)\wedge\tF(k_2)
\rangle_{*_2}}
Eq.\cstaueval\ agrees with the result of Liu and Michelson\refs\LiuQA,
written in Eq.\ifourint, and evaluated at $t=0$ using Eq.\starker. Our
goal is now to reproduce all of Eq.\ifourint, and not just its limit
at $t=0$, from a smearing prescription.

Before proceeding to do this, let us define the abstract symbol $*_2$
as shorthand for the corresponding momentum space kernel:
\eqn\momspacekernel{
*_2 \equiv {\sin{k_1\times k_2\over
2}\over {k_1\times k_2\over 2}} = {\sin\pi a\over \pi a}}
We now modify the smearing prescription by
introducing a weight factor $f(t,\tau)$ into the correlation
function. This amounts to replacing Eq.\csquad\ by:
\eqn\modsmear{
\half\tC^{(6)}(-k)\wedge
\int_0^1 d\tau\,f(t,\tau)\int dx~ F(x)*\wedge\, 
F(x+\theta\cdot k \tau)\,e^{ik.x}
}
The function $f(t,\tau)$ will be determined by requiring agreement
between the effect of this modified smearing, and the amplitude in
Eq.\ifourint. This function should of course reduce to unity in the
Seiberg-Witten limit: $f(t=0,\tau)=1$.

Going to momentum space and dropping the explicit $\tC^{(6)}$, as
before, this amounts to modifying Eq.\cstaueval\ to:
\eqn\cspost{
\half \int_0^1 d\tau\,f(t,\tau)\int dk_1~dk_2~
e^{-{i\over 2}k_1\times k_2}
\tF(k_1)\wedge\tF(k_2)~e^{i(k_1\times k_2)\tau}}
The coefficient of $F\wedge F$ in this expression will become
equivalent to Eq.\ifourint\ if we make the choice:
\eqn\fchoice{
f(t,\tau) = (2\sin\pi\tau)^{2t}}
This follows by writing the $\tau$-integral in Eq.\cspost\ as:
\eqn\ffcoeff{
\half \int_0^1 d\tau\,(2\sin\pi\tau)^{2t}\,
e^{-i\pi a}~e^{2\pi i a\tau}
= \int_0^\half d\tau\,(2\cos\pi\tau)^{2t}\,\cos 2\pi a\tau }
which is precisely the LHS of Eq.\ifourint.

It follows that, to quadratic order in $F$, the complete tree-level
derivative correction to the CS coupling \csquad\ around the SW limit
is given by the insertion of $(2\sin\pi\tau)^{2t}$ in the smearing
prescription.

We would now like to cast this in an equivalent form which will be more
useful. The idea is to find a $t$-dependent deformation of the Moyal
$*$-product in such a way that the full coupling is again given by
\csquad\ but with the new $*$-product replacing the old one. For this,
let us first recall that the kernel which gave rise to the $*_2$
product, on the LHS of Eq.\gammaexp\ is:
\eqn\recallkernel{
{{\Gamma(1+2t)}\over{\Gamma(1+a+t)\Gamma(1-a+t)}} 
~\sim~ {\sin\pi a\over{\pi a}} + {\cal O}(t)}
Now instead of expanding in $t$, we rewrite the exact result as:
\eqn\exactkernel{
{{\Gamma(1+2t)}\over{\Gamma(1+a+t)\Gamma(1-a+t)}} 
= {\sin\pi a\over{\pi a}}\left(
{\Gamma(1+a)\Gamma(1-a)\Gamma(1+2t)\over
\Gamma(1+a+t)\Gamma(1-a+t)}\right) } 
Since the LHS of the above equation reduces to the $*_2$ product when
$t= 0$, it is tempting to think of it, for general $t$, as a deformation
of $*_2$. Thus, generalizing Eq.\momspacekernel, we define:
\eqn\kerdef{
*_2(t)~\equiv~ {{\Gamma(1+2t)}\over{\Gamma(1+a+t)\Gamma(1-a+t)}} }
Notice that $*_2(t)$ is also symmetric under $a \rightarrow -a$ (that
is under $k_1 \leftrightarrow k_2$). It follows that:
\eqn\starck{
*_2(t) = *_2\times \cK(t) }
where
\eqn\ckdef{
\cK(t)~ \equiv~ {\Gamma(1+a)\Gamma(1-a)\Gamma(1+2t)\over
\Gamma(1+a+t)\Gamma(1-a+t)} }
In terms of the deformed $*_2$ product, 
we can write the exact tree-level coupling of $C^{(6)}$ to
two $F$'s, as derived in Ref.\refs\LiuQA, as:
\eqn\coupstart{
\half\tC^{(6)}(-k)\wedge
\int dx~ \langle F\wedge F\rangle_{*_2(t)}\,e^{ik.x} }

By construction, the new kernel $*_2(t)$ has a power series expansion
in $t$, of which the zeroth order term is the usual $*_2$ kernel. Thus
we can define a sequence of kernels via the power series:
\eqn\seqkers{
\eqalign{
*_2(t) &= \sum_{n=0}^\infty *_2^{(n)}~t^n = *_2 + *_2^{(1)}~t +
{\cal O}(t^2) \cr
\cK(t) &= \sum_{n=0}^\infty \cK^{(n)}~t^n = 1 + \cK^{(1)}~t +
{\cal O}(t^2) \cr}}
where 
\eqn\kerzero{
\eqalign{
*_2(t=0) &= *_2^{(0)} = *_2\cr
\cK(t=0) &= \cK^{(0)} = 1\cr}}

The next step is to define a generalized $t$-dependent Moyal
product. Recall that the Moyal $*$-product and $*_2$ kernel are
related via the following equation:
\eqn\startwomoyal{
\int dx\int_0^1 d\tau\, \Big( f(x) * g(x+\theta\cdot
k\tau)\Big)\,e^{ik.x} = \int dx~\langle f,g\rangle_{*_2}\, e^{ik.x} }
This can be generalized in a way that produces the full $*_2(t)$ on
the RHS. Namely, we require a generalization of the Moyal $*$, called
$*(t)$, that satisfies:
\eqn\startwomoyalt{
\int dx\int_0^1 d\tau\, \Big( f(x) *\!(t)~ g(x+\theta\cdot
k\tau)\Big)\,e^{ik.x} = \int dx~\langle f,g\rangle_{*_2(t)}\, e^{ik.x} }
It is easy to check that the solution for $*(t)$ is given by:
\eqn\startsoln{
*\!(t)~\equiv~ * \times \cK(t) }
with $\cK(t)$ given in Eq.\ckdef.

With these definitions, one can write the modified smearing
prescription of Eq.\modsmear\ as an ordinary smearing over the Wilson
line (with no extra function $f(t,\tau)$), but using the 
generalized Moyal product $*(t)$:
\eqn\smearmoyal{
\half\tC^{(6)}(-k)\wedge
\int_0^1 d\tau\,\int dx~ F(x)*\! (t)\wedge\, 
F(x+\theta\cdot k \tau)\,e^{ik.x}
}
By virtue of Eq.\startwomoyalt, this is identical to the desired
result in Eq.\coupstart\ above. Thus we have been led first to a
$t$-dependent $*_2(t)$ product and thence to a Moyal product $*(t)$,
in terms of which the Chern-Simons coupling of two gauge fields to the
RR 6-form can be succinctly expressed. This expression is exact at open
string tree level.

This observation can be followed up in two different directions. On
the one hand, the modified products above will affect the key results
of Ref.\refs{\MukhiVX,\okawao,\LiuPK}, namely the topological identity
and the Seiberg-Witten map obtained from a comparison of commutative
and noncommutative CS couplings. On the other hand, it is important to
check whether the structure of deformed $*$ products remains relevant
when one goes beyond quadratic order in the gauge field. We now
turn to the first of these issues, reserving the second for future
work.

\subsec{Topological Identity Revisited}
In this subsection we reconsider the topological identity of
Refs.\refs{\MukhiVX,\okawao,\LiuPK}. This was originally discovered by
comparing the noncommutative expression for the 10-form coupling of a
D9-brane in the SW limit with the corresponding commutative one. 

We know that in the SW limit $t\to 0$, the noncommutative 10-form
coupling to a D9-brane is: 
\eqn\nctenform{
{\tilde C}^{(10)}(-k)\int dx~
L_*\Big[
\sqrt{\det(1-\theta\hF)}~ W(x,C)
\Big]*e^{ik.x} }
On the other hand, the commutative coupling is:
\eqn\ctenform{
{\tilde C}^{(10)}(-k)\int dx~
e^{ik.x} }
This receives no derivative corrections, as shown in
Ref.\refs\wyllard. Equating the two expressions above
leads to the topological identity:
\eqn\topid{
\int dx~ L_*\Big[{\sqrt{\det(1-\theta \hat F)}}
~ W(x, C)\Big] * e^{ik.x} = \delta(k) }

If we go beyond the SW limit, the commutative 10-form coupling still
receives no corrections, since the argument of Ref.\refs\wyllard\ does
not depend on this limit at all.  On the other hand, the
noncommutative coupling would be changed according to our prescription
above for $t$-deformed $*$-products. Therefore we need to check that
the equivalence of the two couplings continues to hold beyond the SW
limit, as it certainly should.

Let us first review some aspects of the above identity.
For it to hold, all terms in the noncommutative action
must cancel out, to every order in $\hA$, except for the zeroth order term.
Explicit computations upto cubic order in $\hA$ were presented in
Ref.\MukhiVX\foot{A proof of the topological identity in the SW limit
using the matrix model language can be found in
\refs\okawao.}. The cancellation to first order in $\hA$ is very
straightforward, since no smearing is required. Among the quadratic
terms, we have a contribution:
\eqn\amongquad{ 
{i\over 2}\,\theta^{ij}[\hA_j,\hA_i]_* }
coming from the term of first order in $\hF$ from the
Pfaffian, and using the definition:
\eqn\fdef{
\hF_{ij} = \del_i \hA_j - \del_j \hA_i - i[\hA_i,\hA_j]_* }
This term has no smearing, since the smearing prescription
requires us to treat a commutator of gauge fields as a single object. 

Using the identity:
\eqn\famousident{
[f,g]_* = i\,\theta^{kl}\langle \del_k f,\del_l g\rangle_{*_2} }
this term can also be written:
\eqn\famousresult{
\half\,\theta^{ij}\theta^{kl}\langle\del_k\hA_i\,\del_l\hA_j\rangle_{*_2}
} 

There is also a term, coming from quadratic order in the expansion of
the Pfaffian, which is:
\eqn\quadpfaff{
-\half\, \theta^{ij}\theta^{kl}\,\del_k\hA_i(x)*
\del_l\hA_j(x+\theta\cdot k\tau) } 
This term needs to be integrated over $\tau$. As one can check, all
the other terms to quadratic order in $\hA$ cancel each other out
identically {\it even before} $\tau$-integration. Thus, combining
Eqs.\famousresult\ and \quadpfaff, we are left with a
quadratic contribution:
\eqn\quadcont{
\half\,\theta^{ij}\theta^{kl}\int_0^1 d\tau\,\left(
\langle\del_k\hA_i\,\del_l\hA_j\rangle_{*_2} -
\del_k\hA_i(x)* \del_l\hA_j(x+\theta\cdot k\tau)\right) }
to the noncommutative 10-form coupling. 
The first term in this bracket is independent of $\tau$, but it can be
put inside the $\tau$ integral anyway\foot{Also, here and in what
follows we are suppressing the factors $e^{ik.x}$ and $\int dx$.}.
In this form, it is clear that upon integrating over $\tau$, 
the second term gets converted to $*_2$ and cancels the
first. Hence the quadratic terms cancel out as expected.

The question of interest now is what happens to this cancellation when
we go beyond the Seiberg-Witten limit by modifying
the $*$ product. First note that for all the
other quadratic terms, which we have been ignoring because they
cancelled pointwise in $\tau$, the cancellation continues to hold even
if we replace the Moyal $*$ product by $*(t)$. Thus we only need to
concentrate on the terms in Eq.\quadcont\ above, which generalize to:
\eqn\quadcontt{
\half\,\theta^{ij}\theta^{kl}\int_0^1 d\tau\,\left(
\langle\del_k\hA_i\,\del_l\hA_j\rangle_{*_2} -
\del_k\hA_i(x)*\!(t)~ \del_l\hA_j(x+\theta\cdot k\tau)\right) }
Using Eq.\startwomoyalt, this can be rewritten:
\eqn\quadcontre{
\half\,\theta^{ij}\theta^{kl}\,\left(
\langle\del_k\hA_i\,\del_l\hA_j\rangle_{*_2} -
\langle\del_k\hA_i\,\del_l\hA_j\rangle_{*_2(t)} \right) }
and clearly this is no longer zero. 

Now it is clear what must be done in order to restore the cancellation of
quadratic terms. We somehow need to modify the first term above by the
replacement: 
\eqn\firstmod{
\langle\del_k\hA_i\,\del_l\hA_j\rangle_{*_2}
\to \langle\del_k\hA_i\,\del_l\hA_j\rangle_{*_2(t)} }
where $*_2(t)$ is the $t$-deformed $*_2$ kernel defined in
Eq.\kerdef. If this can be done, then the equivalence of commutative
and noncommutative 10-form couplings will be restored. Otherwise the
approach we are proposing will fail to work for the 10-form coupling,
and we will simply not be able to find a general prescription to go
beyond the Seiberg-Witten limit along these lines.

Recall where this first term originated -- it came from the Moyal
commutator in the gauge field strength $\hF$. So the above replacement
can be achieved only if we change the 
definition of the noncommutative field strength $\hF$, to:
\eqn\newfieldstr{
\hF_{ij} = \del_i \hA_j - \del_j \hA_i + 
\theta^{kl}\langle\del_k\hA_i\,\del_l\hA_j\rangle_{*_2(t)} }
In order for the last term to be a commutator of some generalized
product, we need an identity analogous to Eq.\famousident. 
Happily, the required identity
\eqn\famousidentt{
[f,g]_{*(t)} = i\,\theta^{kl}\langle \del_k f,\del_l g\rangle_{*_2(t)} }
can be shown to hold, as a consequence of Eq.\startwomoyalt\ which was
our defining equation for the $t$-deformed Moyal product. It follows
that the correct redefinition of the field strength is:
\eqn\refieldstr{
\hF_{ij} = \del_i \hA_j - \del_j \hA_i -i 
[\hA_i,\hA_j]_{*(t)} }

This amounts to a consistency check. We have learned that upon
replacing the Moyal $*$-product by the expression $*(t)$ {\it both} in
the action {\it and} in the definition of $\hF$, the equivalence of
10-form CS couplings is restored. Another way of saying this is that
the two (in principle independent) ways of defining $*(t)$, via
Eqs.\startwomoyalt\ and \famousidentt\ are equivalent.  Thus our
proposal has passed a crucial test.

This immediately brings up the question of whether our proposal to
deform the $*$ product is in conflict with the theorem of
Kontsevich\refs\kontsevich, which essentially states that the unique
associative, noncommutative product up to certain ``gauge freedoms'',
is the Moyal $*$-product. The answer is that we do not claim $*(t)$ to
be associative. In fact, it is easily checked that $*(t)$ is 
non-associative\foot{Non-associative products generalizing the Moyal $*$
were considered, in a different context, in Ref.\refs\CornalbaSM.}
except in the limit $t\to 0$. We have defined the deformed Moyal
$*$-product only for a pair of functions, since we are working to
quadratic order in $\hA$, so it is not even clear that one should ask
questions about associativity at this stage. When considering higher
order amplitudes, this will clearly be an important issue to
understand.

\subsec{Seiberg-Witten Map Revisited}

We now address the question of how our deformations of $*_2$ and the
Moyal $*$ product affect the Seiberg-Witten map. This map is the
defining equation for a commutative gauge field in terms of a
noncommutative one (or vice-versa). One way of obtaining the SW map is
by comparing the expression for the coupling of a noncommutative
D$p$-brane to the $C^{(p-1)}$ form with its commutative
counterpart\refs{\MukhiVX,\okawao,\LiuPK}. This method relies
crucially on the fact\refs\wyllard\ that one can always choose a basis
of fields on the commutative side, in terms of which the coupling
$\int C^{(p-1)}\wedge F$ does not receive any derivative
corrections. Writing the noncommutative coupling in the SW limit and
equating it to the commutative coupling, one gets:
\eqn\liusidentity{
{\tilde F}_{ij}(k) =
\int dx
~L_*\Big[\sqrt{\det(1-\theta \hat F)}\,
\big({\hat F}(1-\theta{\hat F})^{-1}\big)_{ij}\, W(x, C)\Big] *
e^{ik.x} }
where ${\tilde F}_{ij}(k)$ is the Fourier transform of $F_{ij}$. The
commutative field strength $F_{ij}$ satisfies the Bianchi Identity
\eqn\combi{
\partial_iF_{jk} + \partial_jF_{ki} + \partial_kF_{ij} = 0.}
The RHS of \liusidentity\ is, by construction, invariant under the
noncommutative gauge transformations whose infinitesimal form is given
by:
\eqn\ncgt{
{\hat A}_i \rightarrow {\hat A}_i + \partial_i {\hat \lambda} - i
\big[{\hat A}_i, {\hat \lambda}\big]_{*}. }
Based on our experience with the topological identity in the previous
section, we now propose a modification of the SW map upto terms of
${\cal O}({\hat A}^3)$ which will be valid perturbatively to all
orders in the SW expansion.

Recall that, to lowest order in the SW expansion and to
${\cal O}({\hat A}^2)$, we have:
\eqn\oldswmap{
F_{ij} = {\hat F}_{ij} + \theta^{mn}\langle {\hat F}_{im}, {\hat
F}_{nj}\rangle_{*_2} - {1\over 2} \theta^{mn}\langle {\hat F}_{nm},
{\hat F}_{ij}\rangle_{*_2}
+ \theta^{mn}\partial_n\langle {\hat A}_m,
{\hat F}_{ij}\rangle_{*_2} + {\cal O}({\hat F}^3) }
Consistent with our earlier results, it is natural to propose that
this expression is modified by the replacement of $*_2$ products by
$*_2(t)$ products and by redefining ${\hat F}_{ij}$ to be:
\eqn\newfhat{
{\hat F}_{ij} = \partial_i {\hat A}_j - \partial_j \hat
A_i + \theta^{kl}\langle\partial_k\hat A_i, \partial_l \hat A_j
\rangle_{*_2(t)} .}
As we have seen in the case of topological identity, both these
changes can be achieved by the replacement $* \rightarrow *(t)$ inside
the open Wilson line and in the definition of $\hat F$. With these two
changes, Eq.\oldswmap\ is modified to:
\eqn\simsw{
\eqalign{
F_{ij} = \partial_i{\hat A}_j -\partial_j{\hat A}_i &+
\theta^{mn}\Big[\langle \partial_i {\hat A}_m, \partial_n {\hat A}_j
\rangle_{*_2(t)} + \langle \partial_m {\hat A}_i, \partial_j {\hat A}_n
\rangle_{*_2(t)} - \langle \partial_i {\hat A}_m, \partial_j {\hat A}_n
\rangle_{*_2(t)}\Big] \cr
&+ \theta^{mn}\langle {\hat A}_m, \partial_n(\partial_i{\hat A}_j
- \partial_j{\hat A}_i)\rangle_{*_2(t)} + {\cal O}({\hat A}^3).}}
Notice that this expression for $F_{ij}({\hat A}_i)$ has
exactly the same form as its $t=0$ counterpart except for the change
of $*_2 \rightarrow *_2(t)$. We claim that this is the correct
generalization of the SW map \oldswmap\ to ${\cal O}({\hat A}^2)$ and
perturbatively to all orders in the SW expansion. 

This can be checked in various ways. For example, it can be shown that
this is precisely the expression for the SW map that one obtains by
comparing the commutative and noncommutative couplings of a D$p$-brane
to the RR $C^{(p-1)}$ form in the presence of a constant $B$-field.
We present this argument, which uses the calculations of 
Ref.\refs{\LiuQA}, in the appendix. 

Another check is to show explicitly that the above modification of the
SW map does satisfy the Bianchi identity Eq.\combi. This turns out to
be quite straightforward, so we omit the calculation here. On the other
hand, Eq.\simsw\ is no longer invariant under the gauge
transformations in Eq.\ncgt. To achieve gauge invariance, we find,
perhaps not surprisingly, that we have to also promote the $*$-product
in Eq.\ncgt\ to a $*(t)$-product. Thus the new gauge transformation
law for the noncommutative gauge field ${\hat A}_i$ is:
\eqn\newncgt{
{\hat A}_i \rightarrow {\hat A}_i + \partial_i {\hat \lambda} 
+ \theta^{kl}\langle\partial_k\hat A_i, \partial_l {\hat \lambda}
\rangle_{*_2(t)}.}
Though the validity of the Bianchi identity proves that the LHS of
\simsw\ is a $U(1)$ gauge field, it is nevertheless instructive to 
actually construct explicitly the gauge transformation of the
commutative gauge potential out of the noncommutative gauge
transformation law Eq.\newncgt. Let us first review what is known
about this in the $t=0$ case. A solution to the differential equations
of Ref.\refs\seiwit\ for the SW map to quadratic order can be found in
the paper of Mehen and Wise (the third reference in \refs\starn).
Their solution, upto ${\cal O}({\hat A}^3)$ terms, is:
\eqn\mwone{
A_i = {\hat A}_i + {1\over 2}\theta^{mn}\langle {\hat A}_m,
(2\partial_n {\hat A}_i - \partial_i{\hat A}_n ) \rangle_{*_2} 
+ {\cal O}({\hat A}^3) }
One can check that this is a correct solution by evaluating $F_{ij} =
\partial_i A_j - \partial_j A_i$ and comparing with what one gets from
the open Wilson line prescription (essentially Eq.\simsw\ with $*_2(t)
\rightarrow *_2$). The way Seiberg and Witten set up their equations
ensures, by construction, that the variation of RHS of \mwone\ under:
\eqn\ncgt{
{\hat A}_i \rightarrow {\hat A}_i + \partial_i {\hat \lambda} 
+ \theta^{kl}\langle\partial_k\hat A_i, \partial_l {\hat \lambda}
\rangle_{*_2}}
amounts to the commutative gauge transformation $A_i \rightarrow A_i +
\partial_i \lambda$, provided the two parameters $\lambda$ and ${\hat
\lambda}$ are related to each other by:
\eqn\mwtwo{
\lambda = {\hat \lambda} + {1\over 2}\theta^{mn}\langle{\hat A}_m,
\partial_n{\hat \lambda}\rangle_{*_2} 
+ {\cal O}({\hat A}^2{\hat \lambda}).} 
One can easily check that $\delta_{\hat \lambda}A_i({\hat
A}) = \partial_i \lambda$ using \ncgt, \mwone\ and \mwtwo. For the
sake of completeness, the variation $\delta_{\hat \lambda}A_i({\hat
A})$ of Eq.\mwone\ is:
\eqn\cgt{
\eqalign{
\delta_{\hat \lambda}A_i &= \partial_i {\hat \lambda} + \theta^{mn}
\langle \partial_m {\hat A}_i, \partial_n{\hat \lambda}\rangle_{*_2} 
+ {1\over 2}\big[\langle {\hat A}_m, \partial_n\partial_i {\hat
\lambda} \rangle_{*_2} + \langle \partial_m {\hat \lambda},
(2\partial_n {\hat A}_i - \partial_i {\hat A}_n) \rangle_{*_2} \big]  
+ {\cal O}({\hat A}^2{\hat \lambda})\cr
&= \partial_i \big[{\hat \lambda} + {1\over 2}\theta^{mn} \langle
{\hat A}_m, \partial_n {\hat \lambda}\rangle_{*_2} 
+ {\cal O}({\hat A}^2{\hat \lambda}) \big] = \partial_i \lambda.}}
To go beyond the SW limit, we again replace all $*_2$-products in the
above calculation by $*_2(t)$-products. This actually achieves the
required generalization. Checking this is again trivial, as the
calculation is essentially identical to the one in the $t=0$ case,
with every $*_2$ replaced by $*_2(t)$.

To summarize, the $t\neq 0$ SW map at the level of the gauge
potential, upto ${\cal O}({\hat A}^3)$ terms, is:
\eqn\swmapfora{
A_i = {\hat A}_i + {1\over 2}\theta^{mn}\langle {\hat A}_m,
(2\partial_n {\hat A}_i - \partial_i{\hat A}_n ) \rangle_{*_2(t)} 
+ {\cal O}({\hat A}^3). }
The gauge transformation of the noncommutative gauge field is:
\eqn\newncgt{
{\hat A}_i \rightarrow {\hat A}_i + \partial_i {\hat \lambda} 
+ \theta^{kl}\langle\partial_k\hat A_i, \partial_l {\hat \lambda}
\rangle_{*_2(t)}.}
This generates the commutative gauge transformation 
$A_i \rightarrow A_i + \partial_i \lambda$ provided the two parameters
$\lambda$ and ${\hat \lambda}$ are related by:
\eqn\lhatl{
\lambda = {\hat \lambda} + {1\over 2}\theta^{mn}\langle{\hat A}_m,
\partial_n{\hat \lambda}\rangle_{*_2(t)} 
+ {\cal O}({\hat A}^2{\hat \lambda}).} 

\newsec{Conclusions}

In this paper we have first computed derivative corrections, in the
presence of a constant B-field, to the commutative coupling of a
D9-brane to the RR $C^{(6)}$ form, using boundary-state
techniques. This correction term is of first order beyond the SW
limit, but restricted to quadratic order in gauge fields. Next we
compared it with the expansion of a corresponding world-sheet
calculation of the noncommutative coupling in Ref.\refs\LiuQA, and
found agreement.

Subsequently, we argued that the corrections considered here still
admit an interpretation in terms of straight open Wilson lines on the
noncommutative side. This provides us with a generalization of the
$*_2$ product to all orders in the SW expansion, which we call
$*_2(t)$, where $t$ is essentially the scalar product of the momenta
of the two gauge fields involved. There are two equivalent ways of
defining the $*_2(t)$ product from the point of view of an open Wilson
line. One is to smear the two operators along the length of the Wilson
line using a modified smearing prescription (by inserting a nontrivial
$t$-dependent smearing function). This definition is natural from the
point of view of the open-string world-sheet. The other way is to use
the same smearing prescription as in $t=0$, but to use a deformed
Moyal $*$-product, which we denote by $*(t)$. By looking at the
topological identity, we learned that the definition of the
noncommutative gauge field strength also needs to be changed
correspondingly. We explored the consequences of our observations to
the SW map, again to quadratic order in the gauge fields.

This raises a number of interesting questions. Perhaps the most
important is how to generalize our results beyond quadratic order
in the gauge fields. This involves correctly defining the generalizations
of the higher $*_n$ products. A natural hint comes from the
world-sheet correlators. This suggests that one needs to insert a
factor of $|\sin\pi\tau_{ab}|^{2t_{ab}}$ with $t_{ab}=
\alpha'k_a\cdot k_b$ and $\tau_{ab}=\tau_a-\tau_b$, 
for each pair of operators on the Wilson line at the positions
$\tau_a$ and $\tau_b$. We do not know if this can then be re-expressed
in terms of the $*(t)$ products as well. It will be interesting to
understand the $*(t)$-product better and find a natural place for it
in string theory.

It is known that the noncommutative actions in the SW limit arise
naturally from matrix theory
considerations\refs{\CornalbaAH,\IshibashiVI,\seibnew}. For example
the $*_n$ products of \refs\liu\ arise from the symmetrised trace
prescription of the matrix model action. It is natural to ask what the
deformed $*_2(t)$ product means from the matrix theory
language. Since matrix theory naturally leads to the $\Phi=-B$
description\refs\seibnew, this will require understanding how to
extend our results in Section 5 to general descriptions. This is also
an interesting question in its own right.

Finally, once we go beyond quadratic order in noncommutative gauge
fields, the transcription of these results into commutative gauge
field variables would teach us more about the structure of derivative
corrections to commutative D-brane actions, generalizing the results
in Ref.\refs\dms. We hope to address some of these questions in future
work.

\bigskip\medskip
\noindent{\bf Acknowledgements}

We are happy to acknowledge helpful discussions with Aaron Bergman,
Sergey Cherkis, Aki Hashimoto, Hong Liu, Jeremy Michelson, Juan
Maldacena, Greg Moore, Carlos N\'u\~nez, Nati Seiberg, Ashoke Sen,
Edward Witten and Niclas Wyllard. The work of S.M. is supported in part
by DOE grant DE-FG02-90ER40542 and by the Monell Foundation, while the
work of N.V.S. is supported by a PPARC Research Assistantship.

\appendix{A}{World-Sheet Derivation of \newfhat} 
In Sections (5.1) and (5.2), to satisfy the topological identity and
to have a sensible definition of the SW map we proposed to replace the
$*_2$ products by $*_2(t)$ products and also modify the definition of
the noncommutative field strength ${\hat F}$ as in Eq.\newfhat. In this
appendix, we would like to justify these proposals by looking at the the
1+1 and 1+2--point world-sheet correlation functions on the
disc. Fortunately for us, these amplitudes have been evaluated in
Ref.\refs{\LiuQA}. The details of these calculations can be found in
Section 3 of Ref.\refs{\LiuQA} and will not be reproduced here.

We first note that the correlation functions of one closed-string
vertex operator and two open-string vertex operators (as in
\twooneamp) involve essentially three types of integrals, Eqs. (3.15),
(3.16) and (3.17) of \refs{\LiuQA}. All three integrals can be
evaluated explicitly and are given in Eqs. (D.1a), (D.1b) and (D.1c)
of \refs{\LiuQA}. The proposal of replacing $*_2$ by $*_2(t)$ is
recovered by noticing that all these integrals are proportional to the
$*_2(t)$ kernel given in Eq.\kerdef, up to some simple factors depending
on $k_1$ and $k_2$. Therefore the factor $\big(\sin{k_1\times
k_2 \over 2}\big)/ \big({k_1\times k_2 \over 2}\big)$ in Eq.(3.16) of
\refs{\LiuQA} is actually the $t=0$ part of a full $*_2(t)$
kernel. This proves our first proposal.

Now let us analyze carefully how to recover \newfhat\ from these
amplitudes. For this it is helpful to first trace back to where the
commutator term in ${\hat F}$ at $t=0$ comes from, in Ref.\LiuQA.  We
find ourselves at the last term of Eq.(3.16) of that paper, and the
subsequent discussion where the authors identify this as the term that
combines with the 1+1--point function contribution to give rise to a
noncommutative field strength $f_{MN}$. Now, this term actually came
from the last term of (3.12), the expression for $I_{42}$. This is:
\eqn\ifourtwo{
I_{42} = \cdots + 2 \alpha'^2\, (k_1\cdot k_2){A_2(y) \over
y(1+y^2)}\, a_{1M} a_{2N} \Lambda^{MN}.}
In Ref.\refs{\LiuQA} the amplitude was evaluated only to leading order
in $t$, using (3.15b). But one could use the full answer for this
integral given in (Eq.(D.1b)) of the appendix, which reads:
\eqn\twoint{
\int_{-\infty}^{\infty} dy\, {A_2(y) \over y(1+y^2)} = - {(k_1\times
k_2)\over 2\alpha'k_1\cdot k_2} \times *_2(t)}
Substituting this back into Eq.(3.7) of Ref.\refs\LiuQA\ it is easy to
see that this converts the last term of their Eq.(3.16) into the one
we require. That is, the factor $\big(\sin{k_1\times k_2
\over 2}\big)$ gets replaced by $\big({k_1\times k_2\over 2} \times
*_2(t)\big)$.

One more point to note is the following. There are two different sets
of terms of the kind Eq.\twoint\ from open Wilson lines. One is the
set of commutator terms from field strengths, which are smeared along
the contour of the Wilson line as single objects. The other set of
terms comes from the Wilson line itself, when expanded to quadratic
order in the gauge fields. Now, in the way the terms in the amplitudes
of Ref.\refs\LiuQA\ are written, all the powers of $a_N$ (the
polarization vector of the noncommutative gauge field) that come from
expanding the open Wilson line come with appropriate factors of ${\cal
M}_i$. The one we just looked at is not one of those, and therefore is
genuinely a term that comes from expanding the Pfaffian and other
factors in the open Wilson line which involve the field strengths
rather than potentials. Therefore we conclude that when $\alpha'$
corrections are retained, the definition of $f_{MN}$ in Eq.(3.17) of
\refs{\LiuQA} changes to the one in our Eq.\newfhat. 

Since we have not specified what RR form we are working with (the
$\Lambda$ 's have a formal sum over all RR forms), our conclusion
holds for all such couplings and in particular the couplings to
$C^{(p+1)}$ and $C^{(p-1)}$, which are relevant for the topological
identity and the SW map respectively. This completes the world-sheet
proofs of our generalizations of the topological identity and
redefinition of the SW map to quadratic order.

\listrefs
\end